\documentclass[letterpaper, 10 pt, conference]{ieeeconf}  

\IEEEoverridecommandlockouts                              
\usepackage{epsfig} 
\usepackage{subfig}
\usepackage{graphicx}    
\usepackage{epstopdf} 
\usepackage{amsmath} 
\usepackage{amssymb}  
\usepackage{multirow}
\usepackage[dvipsnames]{xcolor}
\usepackage{soul}
\usepackage[cmintegrals]{newtxmath}
\usepackage{float}
\usepackage{mathrsfs}
\usepackage{algorithmic}
\usepackage{algorithm}
\usepackage{subdepth}

\renewcommand{\ge}{\geq}
\newcommand\e{\mathrm{e}}

\renewcommand{\le}{\leq}

\newcommand\T{\rm T}
\newcommand{\mR}
{\mathbb{R}}

\usepackage{multirow}
\usepackage{color}

\DeclareMathAlphabet\mathbfcal{OMS}{cmsy}{b}{n}

\definecolor{cover}{RGB}{72,0,255}
\definecolor{gold}{RGB}{204,163,0}
\definecolor{darkblue}{RGB}{0,21,120}
\definecolor{rred}{RGB}{225,81,18}
\definecolor{dred}{RGB}{170,60,10}
\definecolor{bblue}{RGB}{0,115,189}
\definecolor{ggreen}{RGB}{0,160,100}
\definecolor{dgreen}{RGB}{0,99,61}
\definecolor{lblue}{RGB}{77,191,237}

\definecolor{d2red}{RGB}{215,48,39}
\definecolor{d2orange}{RGB}{252,141,89}
\definecolor{d2yellow}{RGB}{254,224,144}
\definecolor{d2lblue}{RGB}{224,243,248}
\definecolor{d2blue}{RGB}{145,191,219}
\definecolor{d2dblue}{RGB}{69,117,180}

\newtheorem{Lemma}{Lemma}
\newtheorem{Theorem}{Theorem}

\newtheorem{definition}{Definition}

\newtheorem{Remark}{Remark}
\newtheorem{Assumption}{Assumption}
\newtheorem{Example}{Example}
\newtheorem{Problem}{Problem}

\title{Impact analysis of hidden faults in nonlinear control systems using output-to-output gain}

\author{{Ruslan Seifullaev$^{1}$ and Andr\'{e} M. H. Teixeira$^{1}$}
\thanks{This work has been submitted to IFAC World Congress 2026 for possible publication}
\thanks{\mbox{$^1$}Division of  Systems and Control, Department of Information Technology, Uppsala University, Sweden.}
\thanks{ Email addresses:
        {\tt\footnotesize \{ruslan.seifullaev, andre.teixeira\} @it.uu.se}}
}

\begin{document}

\maketitle
\thispagestyle{empty}
\pagestyle{empty}

\begin{abstract}                
Networked control systems (NCSs) are vulnerable to faults and hidden malfunctions in communication channels that can degrade performance or even destabilize the closed loop. Classical metrics in robust control and fault detection typically treat impact and detectability separately, whereas the output-to-output gain (OOG) provides a unified measure of both. While existing results have been limited to linear systems, this paper extends the OOG framework to nonlinear NCSs with quadratically constrained nonlinearities, considering false-injection attacks that can also manipulate sensor measurements through nonlinear transformations.  Specifically, we provide computationally efficient linear matrix inequality conditions and complementary frequency-domain tests that yield explicit upper bounds on the OOG of this class of nonlinear systems. Furthermore, we derive frequency-domain conditions for absolute stability of closed-loop systems, generalizing the Yakubovich quadratic criterion.
\end{abstract}


\section{Introduction}
Nonlinear networked control systems are increasingly deployed in safety-critical infrastructures such as energy, transportation, and industrial automation. By integrating distributed sensors, actuators, and controllers over shared communication networks, they enable flexible and scalable distributed control with reduced wiring complexity. However, the reliance on wireless and digital communication introduces challenges such as limited bandwidth, latency, packet dropouts, and interference, all of which complicate control design and degrade system performance. Beyond these inherent network-induced limitations, NCSs face additional risks from faults and malicious activities.  Hardware degradation, firmware errors, sensor biases, or adversarial manipulations of communication channels can compromise data integrity and disrupt closed-loop behavior. While certain faults are straightforward to detect, others may remain hidden, subtly altering the system’s nonlinear dynamics without raising alarms. These undetectable faults are particularly critical, as they can gradually degrade performance, reduce robustness, and even lead to instability.

A rich literature exists on the analysis of networked control systems under faults and malicious attacks; see, e.g., \cite{Cardenas08,Sandberg15,22_Teixeira25} and references therein. Among the most frequently studied scenarios is denial-of-service (DoS) \cite{DePersis,23_Dolk17,Seifullaev2024}, where communication between nodes is intermittently disrupted.
Another important class is deception-based attacks \cite{ZHAO2020109128}, in which transmitted data are manipulated to mislead the controller. This class includes several subclasses, such as zero-dynamics attacks \cite{Pasqualetti13}, covert attacks \cite{Smith15}, replay attacks \cite{Mo15}, and other broader families of false-data injection attacks \cite{Liu11}. While many works have focused on fault detection and isolation mechanisms \cite{24_Mousavinejad}, deception-based strategies can remain stealthy and are therefore difficult, if not impossible, to detect. Given these challenges, it is natural to employ quantitative metrics that evaluate both the impact of stealthy attacks and their detectability. 

Classical measures in robust control and fault detection include the $H_\infty$ norm \cite{Zhou96} and the $H_-$ index \cite{WANG07}. However, they have limited applicability to security, as they consider impact and detection separately. To address this limitation, a security metric that combines both performance impact and attack detectability was introduced in \cite{Teixeira2015} (for discrete-time systems) and \cite{Teixeira2021} (for continuous-time systems), termed the output-to-output gain (OOG), which characterizes the adversary's goal of achieving maximum impact while avoiding detection. Nonetheless, existing results on the OOG have been restricted to linear systems, leaving the nonlinear case largely unexplored.

In this paper, we consider nonlinear NCSs subject to malicious faults and attacks. We assume that the adversary can degrade system performance not only by influencing communication channels and injecting additive signals into transmitted measurements, but also by gaining direct access to sensors at the firmware level and altering the sensor readings through a nonlinear transformation (a nonlinear attack). Consequently, the closed-loop system becomes nonlinear, making stability and performance analysis nontrivial. Assuming that the nonlinearities belong to a class characterized by local or integral quadratic constraints, we provide frequency-domain conditions that guarantee absolute stability, based on the Yakubovich quadratic criterion \cite{yakubovich1998,PROSKURNIKOV2015414}. Additionally, we extend the analysis of the OOG to nonlinear systems and, using dissipativity theory, derive a computationally efficient method for estimating this metric.  Therefore, the contributions of this paper can be summarized as follows:
\begin{itemize}
\item {\it Absolute stability analysis under nonlinear faults}. We derive frequency-domain conditions for absolute stability of nonlinear systems with quadratically constrained nonlinearities, generalizing the classical Yakubovich quadratic criterion.
\item
{\it Performance degradation analysis via output-to-output gain}.
We extend the OOG framework to nonlinear systems to quantify the impact of undetectable faults. Using dissipativity theory, we provide computationally efficient linear matrix inequality (LMI) conditions and complementary frequency-domain tests that yield explicit upper bounds on the OOG.
\end{itemize}

The rest of the paper is organized as follows. Section~II presents the detailed problem formulation. In Section~III, we develop the absolute stability analysis based on the Yakubovich quadratic criterion. Section~IV contains the main results on performance analysis using the output-to-output gain. A numerical example illustrating the impact of stealthy nonlinear attacks on system performance is provided in Section~V. Finally, conclusions are drawn in Section~VI.

\section{Problem statement}
Consider the following nonlinear system:
\begin{equation}\label{sys_nonl}
\begin{aligned}
\dot x_{\rm p}(t) &= A_{\rm p}x_{\rm p}(t)+\sum_{i=1}^NQ_{{\rm p},i}\xi_i(t)+B_{\rm p}u(t),\\ 
\sigma_i(t)&=R_{{\rm p},i}x_{\rm p}(t), \quad \xi_i(t)= \varphi_i (\sigma_i(t), t), \,\,\, i=1,\ldots,  N,\\
y_{\rm m}(t) &=C_{\rm m,o}x_{\rm p}(t),\quad y_{\rm p}(t) =C_{\rm p,o}x_{\rm p}(t)+D_{\rm p,o}u(t),
\end{aligned}
\end{equation}
where $x_{\rm p}(t) \in \mR^{n_x}$ is the state vector, $u(t) \in \mR^{n_u}$ is the control input, $y_{\rm m}(t) \in \mR^{n_y}$ is the measurement output, $y_{\rm p}(t) \in \mR^{n_{y_{\rm p}}}$ is the performance output, and $A_{\rm p}$, $B_{\rm p}$, $B_w$, $C_{\rm m,o}$, $C_{\rm p,o}$ and $D_{\rm p,o}$ are the matrices of appropriate dimensions, the pair $(A_{\rm p}, B_{\rm p})$ is controllable and the pair $(A_{\rm p}, C_{\rm m,o})$ is observable. 
The linear part of \eqref{sys_nonl} forms a feedback interconnection with the nonlinear blocks $\varphi_i$, where $\sigma_i(t)\in\mR^{n_{\sigma_i}}$ and $\xi_i(t)\in\mR^{n_{\xi_i}}$ are the input and the output vectors of $\varphi_i$, respectively, and $Q_{{\rm p},i}\in\mR^{n_x\times n_{\xi_i}},  R_{{\rm p},i}\in\mR^{n_x\times n_{\sigma_i}}$ are constant matrices. 

\begin{Assumption}
We assume that the nonlinearities $\varphi_i$ belong to some classes  $\mathfrak{M}_{F_i}=\left\{\left(\xi_i,\sigma_i\right)\right\}$, such that the functions  $\xi_i(t)$ and $\sigma_i(t)$ satisfy the quadratic constraints
\begin{equation}\label{local_constr}
F_i\left(\xi_i(t),\sigma_i(t)\right)\ge 0
\end{equation}
for all $t\ge 0$, where $F_i$ are quadratic forms\footnote{
By a quadratic form $F(z_1,z_2)$ of two vectors, we mean a quadratic form of the concatenated vector  $z=\left[z_1^{\T},z_2^{\T}\right]^{\T}\in \mR^{2n_z}$. Specifically, $F(z_1,z_2) = z^{\T}\bar F z$. This quadratic form can also be extended to the Hermitian form $\tilde F(z)$ as follows: $\tilde F(z)=z^*\bar F z$, where $z\in \mathbb{C}^{2n_z}$.} with the matrices $\bar F_i = \bar F_i^{\T}\in \mR^{\left(n_{\xi_i}+n_{\sigma_i}\right) \times \left(n_{\xi_i}+n_{\sigma_i}\right)}$. 
\end{Assumption}
\begin{figure}
\begin{center}
\includegraphics[scale=0.57]{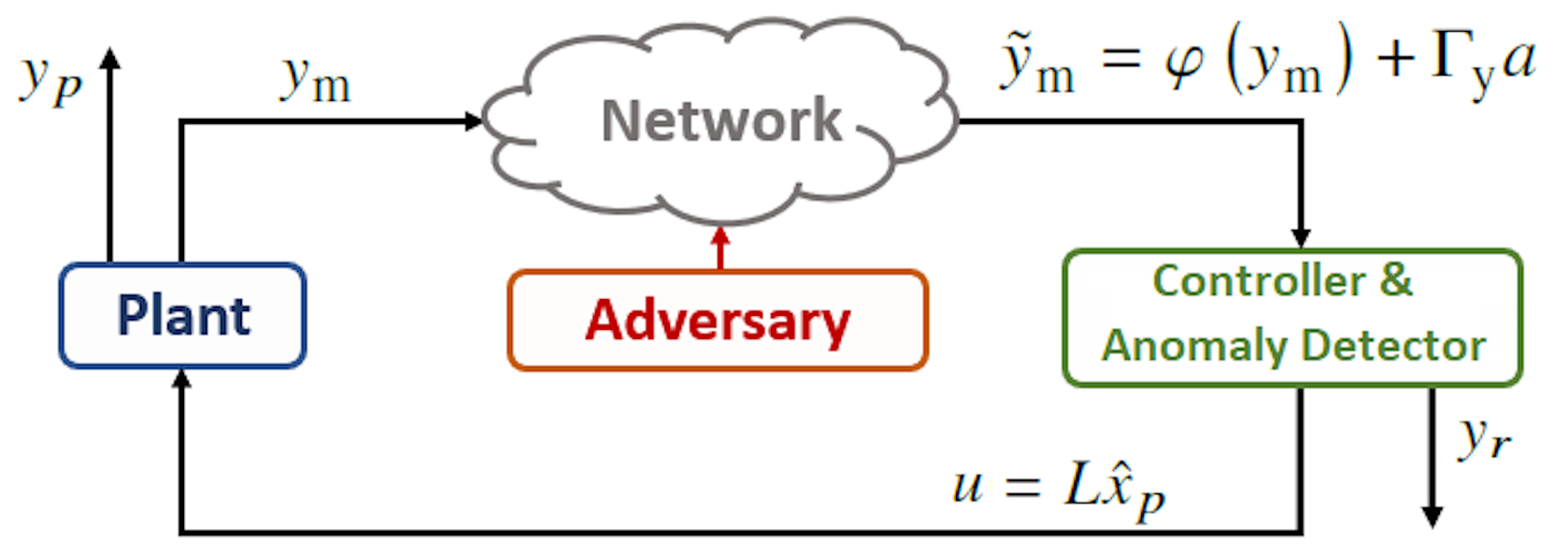}
\end{center}
\caption{The closed-loop system}\label{Nonl_fig}
\end{figure}
\begin{Example}
A typical example of the class  $\mathfrak{M}_F$ is the class  $\mathfrak{M}_{F_{\mu^-, \mu^+}}$ of sector-bounded nonlinearities. Without loss of generality, assume that $n_{\xi}=n_{\sigma}=1$ and $\xi(t)=\varphi\left(\sigma(t),t\right)$ is a nonlinear function satisfying the following sector-bound inequalities:
\begin{equation}\label{eq_sb_ex}
\mu^-\,\sigma\le \xi\le\mu^+\,\sigma,
\end{equation}
for all $\sigma\in\mR$ and $t\ge 0$, where $\mu^-<\mu^+$ are real numbers. 
In this case, the quadratic form $F$ can be written as follows:
$$
F\left(\xi,\sigma\right)=\left(\xi-\mu^-\sigma\right)\left(\mu^+\sigma-\xi\right),
$$
where the inequality $F\left(\xi,\sigma\right)\ge0$ is equivalent to \eqref{eq_sb_ex}.
\end{Example}
\begin{Remark}
We can also consider the uncertain system
$$
\dot x_{\rm p}(t) = (A_{\rm p}+\Delta A_{\rm p})x_{\rm p}(t)+\sum_{i=1}^{N_1}Q_{{\rm p},i}\xi_i(t)+B_{\rm p}u(t),
$$
where the components of the unknown matrix  $\Delta A_{\rm p}$ are bounded. In this case, the system can be represented in form \eqref{sys_nonl}, where the uncertainties, after some structural changes,  can be expressed in terms of nonlinearities $\varphi_j$, $j=N_1+1, \ldots, N_2$, with $N=N_1+N_2$, see \cite{IJRNC15} for details.   
\end{Remark}

\begin{figure}
\begin{center}
\includegraphics[scale=0.25]{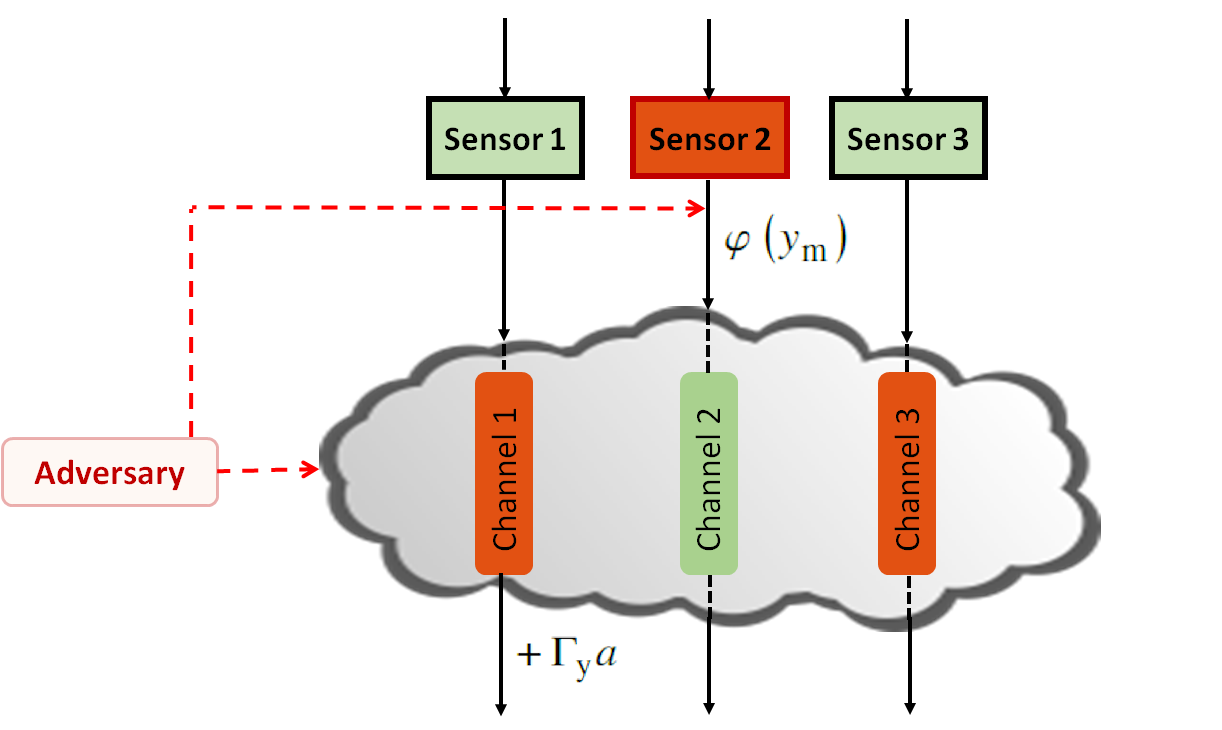}
\end{center}
\caption{Nonlinear and additive attacks}\label{Nonl_fig2}
\end{figure}

We consider the scenario in which an adversary can access the sensor measurements and replace the vector $y_{\rm m}(t)$ with 
\begin{equation}\label{nonl_att_output_additive}
\tilde y_{\rm m}(t) = \varphi\!\left(y_{\rm m}(t),t\right) + \Gamma_{\rm y} a(t),
\end{equation}
where $a(t)\in\mR^{n_a}$ and $\Gamma_{\rm y}\in\mR^{n_y\times n_a}$.
Therefore, instead of the true measurement output $ y_{\rm m}$, the controller receives the modified signal $\tilde y_{\rm m}$, see Figure~\ref{Nonl_fig}.
\begin{Assumption}\label{as2}
In analogy with the system nonlinearities, the nonlinear transformation $\varphi\!\left(y_{\rm m}(t),t\right)$ is taken from a class $\mathfrak{M}_{F_{\rm m}}$ characterized by the quadratic  form
\begin{equation}\label{local_constr_att}
F_{\rm m}\!\left(\xi_{\rm m}(t),y_{\rm m}(t)\right) \ge 0
\end{equation}
with the matrix $\bar F_{\rm m} = \bar F_{\rm m}^{\T} \in \mR^{2n_y \times 2n_y}$.
\end{Assumption}
\begin{Assumption}\label{as3}
The signal $a(t)$ is assumed to lie in the extended $L_2$ space, defined as
$$
L_{2e} = \left\{ a: \mR_+ \to \mR^{n_a} \,\big|\, \|a\|_{L_{2[0,T]}} < \infty,\ \forall T < \infty \right\},
$$
which represents all signals that are square integrable over finite periods of time.
\end{Assumption}
\begin{Remark}\label{Rem2}
The matrix $\Gamma_{\rm y}$ specifies the channels into which the additive signal $a(t)$ can be injected. If $\Gamma_{\rm y}$ is the identity matrix, the adversary can influence all channels and arbitrarily modify all transmitted signals. In this case, considering the nonlinear term in \eqref{nonl_att_output_additive} is redundant, since it can be fully absorbed into $a(t)$. However, injecting a continuous signal into all channels may be energy-intensive and practically difficult or even infeasible. In practice, typically only a subset of channels can be affected in real time. Nonetheless, the adversary may also gain direct access to the sensors and alter the remaining channel signals at the firmware level through a (usually stationary) nonlinear transformation $\varphi$. This makes scenarios involving nonlinear attacks more challenging than those with purely additive attacks, see Figure~\ref{Nonl_fig2}.
\end{Remark}
\begin{Remark}\label{Rem3}
Assumption~\ref{as2} is reasonable, as an adversary typically seeks to degrade system performance without causing complete instability; quadratically constrained nonlinearities effectively capture this behavior. 
Note that, similarly to  $\Gamma_{\rm y}$, the constraint  \eqref{local_constr_att} can specify which sensors are affected or not by a transformation. For example, if some components $\xi_{{\rm m},j} = y_{{\rm m},j}$ for certain indices $j$, then the corresponding block in $\bar F_{\rm m}$ is $\begin{bmatrix}-1&1\\1&-1 \end{bmatrix}$, which is equivalent to setting $\mu^-=\mu^+=1$ in \eqref{eq_sb_ex}. 

Assumption~\ref{as3} is also not restrictive, as the space $L_{2e}$ represents most well-behaved signals encountered in practice.
\end{Remark}

\subsection{Observer-based control and anomaly detector}
To estimate the state vector $x_{\rm p}(t)$ and design a feedback based on it, we use an observer-based state feedback controller. Additionally, we use its residual output $y_{\rm r}(t)$, the difference between the measured and predicted outputs,  for an anomaly detector, which raises an alarm if the residual becomes abnormally high.

We consider the following controller structure: 
\begin{equation}\label{obs_lin}
\begin{aligned}
\dot{\hat x}_{\rm p}(t) &= A_{\rm p}\hat x_{\rm p}(t)+B_{\rm p}u(t)+Ky_{\rm r}(t),\,\,\,\,
u(t) = L\hat x_{\rm p}(t),\\
\hat y_{\rm m}(t) &=C_{\rm m,o}\hat x_{\rm p}(t),\,\,\,\,
y_{\rm r}(t) = \tilde y_{\rm m}(t)-\hat y_{\rm m}(t),
\end{aligned}
\end{equation}
where $K\in\mR^{n_x\times n_y}$ and $L\in\mR^{n_u\times n_x}$ are the observer and controller gains, respectively.
Define $
x=  \begin{bmatrix}
x_{\rm p}\\
x_{\rm p}-\hat x_{\rm p}\\
\end{bmatrix}.
$
Then the resulting closed-loop system can be written as follows:
\begin{equation}\label{sys_nonl_cl}
\begin{aligned}
\dot x(t) &= Ax(t)+\sum_{i=1}^NQ_i\xi_i(t)+Q_{\rm m}\xi_{\rm m}(t)+Ba(t),\\ 
\sigma_i(t) &= R_ix(t), \quad \xi_i(t)= \varphi_i (\sigma_i(t), t), \quad i=1,\ldots,  N,\\
y_{\rm m}(t) &= C_{\rm m}x(t), \quad \xi_{\rm m}(t)=\varphi (y_{\rm m}(t), t),  \\
y_{\rm p}(t) &= C_{\rm p}x(t), \quad y_{\rm r}(t) =C_{\rm r}x(t)+\xi_{\rm m}(t)+D_{\rm r}a(t),
\end{aligned}
\end{equation}
where
\begin{equation*}
\begin{aligned}
 A & = \begin{bmatrix}
A_{\rm p}+B_{\rm p}L&-B_{\rm p}L\\
KC_{\rm m,o}&A_{\rm p}-KC_{\rm m,o}\\
\end{bmatrix},\,\, 
Q_i=\begin{bmatrix}
Q_{{\rm p},i}\\
Q_{{\rm p},i}\\
\end{bmatrix},
\,\, 
Q_{\rm m}=\begin{bmatrix}
0\\
-K\\
\end{bmatrix},\\
B &= \begin{bmatrix}
0\\
-K\Gamma_{\rm y}\\
\end{bmatrix}, \,\,
R_i   = \left[ 
R_{{\rm p},i},\,\, 0
\right], \,\, C_{\rm m}  = \left[ 
C_{\rm m,o},\,\, 0
\right], \,\,D_{\rm r}=\Gamma_{\rm y},\\
C_{\rm p}  &= \left[ 
C_{\rm p,o}+D_{\rm p,o}L,\, -D_{\rm p,o}L
\right], \,\,C_{\rm r}  = \left[ 
-C_{\rm m,o},\,\, C_{\rm m,o}
\right].
\end{aligned}
\end{equation*}

\begin{Remark}[Anomaly detector] In this paper, we do not address the specific design of anomaly detectors; however, a typical anomaly detector raises an alarm if the energy of $y_{\rm r}$ exceeds a certain threshold, i.e.,
\begin{equation}\label{detector}
||y_{\rm r}||^2_{L_2}> \varepsilon_{\rm tr}.
\end{equation}
\end{Remark}
Thus, it is reasonable for the adversary to aim at degrading system performance without triggering detection.

\subsection{Problem formulation}
As discussed in Remarks~\ref{Rem2} and \ref{Rem3}, the nonlinear transformation can be applied by the adversary at the sensor firmware level independently of the additive term $a(t)$. Therefore, it is reasonable to assume that the class $\mathfrak{M}_{F_{\rm m}}$ is chosen so that closed-loop system stability is preserved. However, in the presence of quadratic nonlinearities, deriving a closed-loop stability criterion is more challenging than in the linear case. Consequently, we first address the following problem:
\begin{Problem}\label{Prblm1}
For $a(t)\equiv0$, determine the conditions guaranteeing absolute stability of \eqref{sys_nonl_cl}. 
\end{Problem}

Next, we consider the performance problem, in which the adversary chooses the additive signal $a(t)$ to degrade system performance, as measured by $||y_{\rm p}||_{L_2}$, while simultaneously avoiding complete instability and remaining undetected, i.e., keeping $||y_{\rm r}||_{L_2}$ small. 
\begin{Problem}\label{Prblm2} For the closed-loop nonlinear system \eqref{sys_nonl_cl}, characterize the worst-case impact of the undetected attack \eqref{nonl_att_output_additive}, specifically when the energy of the performance output $y_{\rm p}$ is maximized while the energy of the residual output $y_{\rm r}$ remains small.
\end{Problem}

\section{Absolute stability analysis}
In this section, we investigate Problem~\ref{Prblm1} and introduce the absolute stability criterion for the closed-loop system \eqref{sys_nonl_cl} with $a(t)\equiv0$. For simplicity, we begin with a scenario where the system \eqref{sys_nonl} is linear, i.e., $N=0$. In this case, the closed-loop system \eqref{sys_nonl_cl} has only one nonlinearity, $\xi_{\rm m}$. We then extend this analysis to general nonlinear systems.

Consider the  closed-loop system \eqref{sys_nonl_cl} with $a(t)\equiv0$ and $N=0$:
\begin{equation}\label{sys_lin}
\begin{aligned}
\dot x(t) &= Ax(t)+Q_{\rm m}\xi_{\rm m}(t),\\ 
y_{\rm m}(t) &= C_{\rm m}x(t), \quad \xi_{\rm m}(t)=\varphi (y_{\rm m}(t), t),
\end{aligned}
\end{equation}
where the nonlinearity  $\varphi\!\left(y_{\rm m}(t),t\right)$ belongs to the class $\mathfrak{M}_{F_{\rm m}}$ from Assumption~\ref{as2}.
\begin{Remark}
When considering sector-bounded nonlinearities,  i.e., $\mathfrak{M}_{F_{\rm m}}=\mathfrak{M}_{F_{\mu^-, \mu^+}}$ from Example~1, a natural question arises: if the system is stable for all linear functions $\varphi(y_{\rm m})=\mu y_{\rm m}$ with $\mu\in[\mu^-, \mu+]$, will the closed-loop system remain stable for arbitrary nonlinearities from the same sector? This is the well-known Aizerman conjecture \cite{Aizerman49,leonov2001}, and has since been refuted by many researchers. 
A sufficient condition of absolute stability in the class $\mathfrak{M}_{F_{\mu^-, \mu^+}}$ is given by the circle criterion \cite{yakubovich2004,Khalil}:
Assume that 
\begin{enumerate}
\item the transfer matrix $G_{\rm m}(s)=C_{\rm m}(sI-A)^{-1}Q_{\rm m}$ has no poles on the imaginary axis;
\item the closed-loop system with $\varphi(y_{\rm m})=\mu y_{\rm m}$ is stable for some $\mu\in[\mu^-, \mu^+]$;
\item the following frequency condition holds for all $\omega\in\mR$:
\begin{equation}\label{Circle_freq}
{\rm Re}\left\{[1+\mu^-G_{\rm m}(i\omega)][1+\mu^+G_{\rm m}(i\omega)]^*\right\}>0.
\end{equation}
\end{enumerate}
Then the closed-loop system  \eqref{sys_lin}, \eqref{eq_sb_ex} is exponentially stable. Note that the condition \eqref{Circle_freq} is equivalent to requiring that the Nyquist curve $G_{\rm m}(i\omega)$ does not enter or enclose the circle defined by the points $-1/\mu^-$ and $-1/\mu^+$.
\end{Remark}

Next, we provide absolute stability conditions for a general class $\mathfrak{M}_{F_{\rm m}}$ satisfying \eqref{local_constr_att}.
\begin{definition}
The closed-loop system \eqref{sys_lin} is called {\it absolutely stable} in the class $\mathfrak{M}_{F_{\rm m}}$ if for any solution $x(t)$ satisfying $\left(\xi_{\rm m}(t),y_{\rm m}(t)\right)\in\mathfrak{M}_{F_{\rm m}}$ 
there exists a constant $c_1>0$ such that
\begin{equation}\label{abs_stab_eq1}
\Vert x\Vert^2_{L_2}+\Vert \xi_{\rm m}\Vert^2_{L_2}\le c_1 \Vert x(0)\Vert^2.
\end{equation}
\end{definition}
\begin{definition}
The closed-loop system \eqref{sys_lin} is called {\it minimally stable} in the class $\mathfrak{M}_{F_{\rm m}}$ if there exist a solution $x(t)$ satisfying $\left(\xi_{\rm m}(t),y_{\rm m}(t)\right)\in\mathfrak{M}_{F_{\rm m}}$ 
such that
\begin{equation*}\label{min_stab_eq}
\lim_{t\to\infty}\Vert x(t)\Vert=0.
\end{equation*} 
\end{definition}

The following theorem, \cite{yakubovich1998}, provides a sufficient condition for absolute stability.
\begin{Theorem}[The Yakubovich quadratic criterion]\label{Prop_Quadr}
Assume that the matrix $G_{\rm m}(s)$ has no poles on the imaginary axis and the closed-loop system \eqref{sys_lin} is minimally stable in the class $\mathfrak{M}_{F_{\rm m}}$. Then it is absolutely stable if the following frequency condition\footnote{An obvious advantage of frequency methods is the conceptual simplicity of calculating frequency responses when mathematical models are available. Moreover, even in situations where mathematical models are absent, frequency responses can often be obtained experimentally.} is satisfied:
\begin{equation}\label{freq_quadr_crit}
\tilde F_{\rm m}(i\omega, \tilde \xi_{\rm m})<0,\quad \mbox{for all} \,\, \omega\in \mR \,\,\, \mbox{and}\,\, \tilde\xi_{\rm m}\in \mathbb{C}^{n_y}, \tilde\xi_{\rm m}\ne 0,
\end{equation}
where the Hermitian form $\tilde F_{\rm m}$ is the extension of $F_{\rm m}$ obtained as follows:
$$
\tilde F_{\rm m}\left(s,\tilde \xi_{\rm m}\right) =  \begin{bmatrix}
\tilde \xi_{\rm m}\\
G_{\rm m}(s)\tilde \xi_{\rm m}\\
\end{bmatrix}^*\bar F_{\rm m} \,\begin{bmatrix}
\tilde \xi_{\rm m}\\
G_{\rm m}(s)\tilde \xi_{\rm m}\\
\end{bmatrix}.
$$
Moreover, the absolute stability is exponential: there exist constants $c>0$ and $\alpha>0$ such that
$$
\Vert x(t)\Vert^2\le c \e^{-2\alpha t}\Vert x(0)\Vert^2, \quad t\ge 0.
$$
\end{Theorem}
\begin{Remark}
In the Circle criterion, Condition (2) corresponds to the minimal stability of the closed-loop system in the class $\mathfrak{M}_{F_{\mu^-, \mu^+}}$. Condition (3) is obtained from \eqref{freq_quadr_crit} with 
$$
\begin{aligned}
\tilde F\!\left(i\omega,\tilde \xi\right) &= \begin{bmatrix}
\tilde \xi_{\rm m}\\
G_{\rm m}(i\omega)\tilde \xi_{\rm m}\\
\end{bmatrix}^*\!
\begin{bmatrix}
-1& \frac{1}{2}(\mu^-+\mu^+)\\ *& -\mu^-\mu^+\\
\end{bmatrix}
\begin{bmatrix}
\tilde \xi_{\rm m}\\
 G_{\rm m}(i\omega)\tilde \xi_{\rm m}\\
\end{bmatrix}\\
&=-\left|\tilde\xi_{\rm m}\right|^2{\rm Re}\left\{[1+\mu^-G_{\rm m}(i\omega)][1+\mu^+G_{\rm m}(i\omega)]^*\right\}\!.
\end{aligned}
$$ 
\end{Remark}
\begin{Remark}
The quadratic criterion in Theorem~\ref{Prop_Quadr} also applies to integral quadratic constraints (IQCs), where the class $\mathfrak{M}_{F_{\rm m}}$ is such that the functions  $\xi_{\rm m}(t)$ and $y_{\rm m}(t)$ satisfy
\begin{equation}\label{iqc}
\int_0^{t_k}F_{\rm m}\left(\xi_{\rm m}(t),y_{\rm m}(t)\right)dt\ge 0,
\end{equation}
for some sequence $t_k\to\infty$ as $k\to\infty$. 
Moreover, in the case of IQCs, the condition \eqref{freq_quadr_crit} is also necessary for absolute stability, \cite{yakubovich1998}.  
\end{Remark}

Next, we generalize the Yakubovich quadratic criterion to nonlinear systems with $N>0$ in \eqref{sys_nonl_cl}.
Define the augmented vectors and matrices
\begin{equation}\label{form_augm_not}
\begin{aligned}
\xi(t) &= \begin{bmatrix}
\xi_1(t)\\
\vdots\\
\xi_N(t)\\
\xi_{\rm m}(t)\\
\end{bmatrix},
\,\,
\sigma(t)=\begin{bmatrix}
\sigma_1(t)\\
\vdots\\
\sigma_N(t)\\
y_{\rm m}(t)\\
\end{bmatrix},
\,\, 
R=\begin{bmatrix}
R_1\\
\vdots\\
R_N\\
C_{\rm m}\\
\end{bmatrix},\\
Q &= \left[Q_1, \ldots, Q_N, Q_{\rm m}\right], \quad Q_{\rm r} = \left[0, \ldots, 0, I\right],
\end{aligned}
\end{equation}
and the quadratic form
\begin{equation}\label{form_augm}
F(\xi,\sigma)=\sum_{i=1}^N\tau_iF_i(\xi_i,\sigma_i)+\tau_{\rm m}F_{\rm m}(\xi_{\rm m},y_{\rm m}),
\end{equation}
where $\tau_i>0$, $\tau_{\rm m}>0$ are real scalars.
The following theorem extends the absolute stability criterion from Theorem~\ref{Prop_Quadr} to the closed-loop system \eqref{sys_nonl_cl} and is the solution to Problem~1.
\begin{Theorem}\label{Thm_Quadr_nonl}
Assume that $a(t)\equiv0$, the matrix $G_{\!\xi\!\sigma}(s)=R(sI-A)^{-1}Q$ has no poles on the imaginary axis, and the closed-loop system \eqref{sys_nonl_cl} is minimally stable in the class $\mathfrak{M}_F$ with the form $F$ is defined by \eqref{form_augm}. Then it is absolutely stable if \eqref{freq_quadr_crit} holds with
$$\tilde F(s, \tilde\xi)=F(\tilde\xi,G_{\!\xi\!\sigma}(s)\tilde\xi).$$
\end{Theorem}
\begin{proof}
The state equation in \eqref{sys_nonl_cl} can be rewritten as
$$
\dot x(t)=Ax(t)+Q\xi(t), \,\,\,
\sigma(t) = Rx(t), \,\,\, \xi(t)= \Phi (\sigma(t), t),
$$
where
\begin{equation}\label{Phi_Def}
\Phi(\sigma(t),t) = \left[
\varphi^{\T}_1(\sigma_1(t),t),
\ldots,
\varphi^{\T}_N(\sigma_N(t),t),
\varphi^{\T}(y_{\rm m}(t),t)\right]^{\T}\!\!\!.
\end{equation}
Since all $\varphi_i$ and $\varphi$ satisfy \eqref{local_constr} (or \eqref{iqc}) and \eqref{local_constr_att} (or \eqref{iqc}), respectively, it follows from \eqref{form_augm} that the nonlinearity $\Phi$ belongs to the class $\mathfrak{M}_F$ with the form $F(\xi,\sigma)$ satisfying \eqref{local_constr} (or \eqref{iqc}). 
The result then follows directly from Theorem~\ref{Prop_Quadr}.
\end{proof}
\begin{Remark}
For nonlinear attacks, the anomaly detector \eqref{detector} can be augmented with the condition
the condition
\begin{equation}\label{detector_nonl}
F\left(\tilde y_{\rm m}(t),\hat y_{\rm m}(t)\right)< \varepsilon,
\end{equation}
where the nonlinearity approaches the boundary of the class  $\mathfrak{M}_F$. 
The absence of an alarm would indicate that the assumptions on $\Phi$ remain satisfied, and thus all results continue to hold.
\end{Remark}

\section{Main result: performance analysis using the output-to-output gain}
As previously discussed, the adversary's goal is to degrade the performance, measured by $||y_{\rm p}||_{L_2}$, while simultaneously avoiding making the system completely unstable and remaining undetected, i.e., keeping $||y_{\rm r}||_{L_2}$ small. 
This type of attack is known as a stealthy attack.
The adversary can achieve this by applying the nonlinear transformation $\varphi$ to the measured output $y_{\rm m}(t)$ while also injecting an additive signal $\Delta y=\Gamma_{\rm y} a(t)$ to the sensors' output, as it is shown in \eqref{nonl_att_output_additive}.

A security metric that combines both performance impact and attack detectability was introduced in \cite{Teixeira2015,Teixeira2021}, termed the output-to-output gain (OOG). 
The OOG metric is formulated as the optimal control problem:
\begin{equation}\label{OOG}
OOG\triangleq  \sup\limits_{a\in L_{2e}, x(0)=0}||y_{\rm p}||^2_{L_2}, \quad \mbox{s.t.}\quad ||y_{\rm r}||^2_{L_2}\le1.
\end{equation}
In other words, the OOG characterizes the adversary's goal of achieving maximum impact while avoiding detection.
\begin{definition}[\cite{Khalil}]
The closed-loop system \eqref{sys_nonl_cl} is said to be {\it output strictly dissipative} with respect to a supply rate $w(x(t),a(t))$ if there exists a continuously differentiable positive semidefinite storage function $V(x(t))$ such that $V(0)=0$ and
\begin{equation}\label{def_dissipativity}
\dot V(x(t)) \le w(x(t),a(t))-{\gamma} y_{\rm p}^{\T}(t)y_{\rm p}(t), \quad \gamma>0.
\end{equation}
\end{definition}
\begin{Lemma}\label{Lem_dissipativity}
Assume that \eqref{sys_nonl_cl} is output strictly dissipative with respect to the supply rate $w(x(t),a(t))=y_{\rm r}^{\T}(t)y_{\rm r}(t)$. Then it is finite-gain $L_2$ output-to-output stable, i.e.,
\begin{equation}\label{OOStab}
||y_{\rm p}||^2_{L_2}\le\frac{1}{\gamma}||y_{\rm r}||^2_{L_2}+\beta(x(0)),
\end{equation}
where $\beta(x(0))={\rm const}\ge 0$. In addition,
\begin{equation*}
OOG\le\frac{1}{\gamma}.
\end{equation*}
\end{Lemma} 
\begin{proof}
From \eqref{def_dissipativity}, we obtain
\begin{equation}\label{pf_dissipativity}
\dot V(x(t)) \le y_{\rm r}^{\T}(t)y_{\rm r}(t)-{\gamma} y_{\rm p}^{\T}(t)y_{\rm p}(t).
\end{equation}
Integrating both sides of \eqref{pf_dissipativity}, we get
$$
\begin{aligned}
&\int_0^\tau y_{\rm p}^{\T}(t)y_{\rm p}(t)dt\le\frac{1}{\gamma}\int_0^\tau y_{\rm r}^{\T}(t)y_{\rm r}(t)dt\\
&-\frac{1}{\gamma} \left(V(x(\tau))-V(x(0)\right)
\le\frac{1}{\gamma}\int_0^\tau y_{\rm r}^{\T}(t)y_{\rm r}(t)dt+\frac{1}{\gamma} V(x(0)).
\end{aligned}
$$
For $\tau\to\infty$, the latter implies \eqref{OOStab} with $\beta(x(0))=\frac{1}{\gamma} V(x(0))$.
Finally, for $x(0)=0$ and $||y_{\rm r}||^2_{L_2}\le1$, we obtain $||y_{\rm p}||^2_{L_2}\le\frac{1}{\gamma}$,
which makes $\frac{1}{\gamma}$ the upper bound of the $OOG$.
\end{proof}

Assume that the class $\mathfrak{M}_F$ is defined as a set of pairs $\left\{(\xi,\sigma)\right\}$ satisfying the quadratic constraint
\begin{equation}\label{local_constr_thm}
F\left(\xi,\sigma\right)=\begin{bmatrix}
\xi\\
\sigma\\
\end{bmatrix}^{\T}\!
\bar F
\begin{bmatrix}
 \xi\\
\sigma\\
\end{bmatrix}\ge 0,
\end{equation}
where the matrix $\bar F$ is a symmetric matrix with the following block structure:
$$
\bar F=\begin{bmatrix}F_{11}& F_{12}\\ * & F_{22}\\
\end{bmatrix}.
$$
Consider the matrices
$$
\begin{aligned}
\Psi_0(P) & = \begin{bmatrix}
A^{\T}P+PA& PQ & PB\\
* & 0 & 0\\
* & * & 0 \end{bmatrix}, \\ 
\Psi_1(\gamma) &= \begin{bmatrix}
{\gamma} C_{\rm p}^{\T}C_{\rm p}-C_{\rm r}^{\T}C_{\rm r} & -C_{\rm r}^{\T}Q_{\rm r} & -C_{\rm r}^{\T}D_{\rm r}\\
* & -Q_{\rm r}^{\T}Q_{\rm r} & -Q_{\rm r}^{\T}D_{\rm r}\\
* & * & -D_{\rm r}^{\T}D_{\rm r} \end{bmatrix},\\
\Psi_2(\kappa) &= \kappa \begin{bmatrix}
r^{\T}F_{22}r& r^{\T}F_{12}^{\T} & 0\\
* & F_{11} & 0\\
* & * & 0 \end{bmatrix}.
\end{aligned}
$$
The following theorem solves Problem~2 and provides a computational approach to obtain the upper bound of the $OOG$ the nonlinear system \eqref{sys_nonl} under the sensor measurement attack in \eqref{nonl_att_output_additive}.
\begin{Theorem}\label{LMI_OOG}
Consider the closed-loop nonlinear system \eqref{sys_nonl_cl}.
Assume that there exist a matrix $P\ge0$ and scalars $\kappa\ge0$ and $\gamma>0$ such that the LMI
\begin{equation}\label{thm_LMI}
\Psi_0(P)+\Psi_1(\gamma)+\Psi_2(\kappa)\le0
\end{equation}
is feasible. Then 
\begin{equation}\label{OOG_bound}
OOG\le\frac{1}{\gamma}.
\end{equation}
\end{Theorem}
\begin{proof}
Consider the function $V(x)=x^{\T}P{x}$.
Then
\begin{equation}\label{dotV_thm}
\begin{aligned}
&\dot V-y_{\rm r}^{\T}y_{\rm r}+{\gamma} y_{\rm p}^{\T}y_{\rm p}
=x^{\T}P(Ax+Q\xi+Ba)\\
&+\left(x^{\T}A^{\T}+\xi^{\T}Q^{\T}+a^{\T}B^{\T}\right)Px +{\gamma} x^{\T}C_{\rm p}^{\T}C_{\rm p}x\\
&-\left(x^{\T}C_{\rm r}^{\T}+\xi^{\T}Q_{\rm r}^{\T}+a^{\T}D_{\rm r}^{\T}\right)\left(C_{\rm r}x+Q_{\rm r}\xi+D_{\rm r}a\right)=G(\eta),
\end{aligned}
\end{equation}
where $\eta=\left[x^{\T},\xi^{\T},a^{\T}\right]^{\T}$ and 
$
G(\eta) = \eta^T \Psi_0(P)\eta + \eta^T \Psi_1(\gamma)\eta.
$
For dissipativity of \eqref{sys_nonl_cl}, the quadratic form $G(\eta)$ must be nonpositive.
Taking into account the constraint \eqref{local_constr_thm}, we obtain the following problem: $G(\eta)\le0$ for  $\eta$ satisfying the constraint \eqref{local_constr_thm}.
This problem can be rewritten using using S-procedure, \cite{Yak71}: there exists $\kappa\ge0$ such that
$$
G(\eta)+ \eta^T \Psi_2(\kappa)\eta\le0, \quad \forall \eta,
$$
which coincides with \eqref{thm_LMI}. 
Therefore, the dissipativity of \eqref{sys_nonl_cl}, \eqref{local_constr_thm} follows from the feasibility of the LMI \eqref{thm_LMI}. 
Then, we immediately conclude \eqref{OOG_bound} from Lemma~\ref{Lem_dissipativity}. 
\end{proof}
\begin{Remark}Since the $OOG$ is bounded by $\frac{1}{\gamma}$, the best estimate can be obtained by finding the maximum $\gamma>0$ for which the LMI \eqref{thm_LMI} is feasible.
\end{Remark}

Next, we will provide frequency domain conditions guaranteeing the upper bound for the output-to-output gain.
Introduce the following transfer matrices:
$$
\begin{aligned}
G_{\!a{\rm p}}(s) &=C_{\rm p}(sI-A)^{-1}B,\quad
G_{\!a{\rm r}}(s) =C_{\rm r}(sI-A)^{-1}B+D_r,\\
G_{\!\xi{\rm p}}(s) &=C_{\rm p}(sI-A)^{-1}Q,\quad
G_{\!\xi{\rm r}}(s) =C_{\rm r}(sI-A)^{-1}Q+Q_{\rm r},\\
G_{\!a\!\sigma}(s) &=R(sI-A)^{-1}B, \quad\,
G_{\!\xi\!\sigma}(s) =R(sI-A)^{-1}Q.
\end{aligned}
$$ 
Define also the matrix
$$
\Psi(s,\gamma,\kappa)=\begin{bmatrix}
\Psi_{11}(s,\gamma,\kappa) & \Psi_{12}(s,\gamma,\kappa)\\
* & \Psi_{22}(s,\gamma,\kappa)\end{bmatrix},
$$
where
$$
\begin{aligned}
\Psi_{11}(s,\gamma,\kappa) &= {\gamma}G_{\!\xi{\rm p}}^*(s)G_{\!\xi{\rm p}}(s)-G_{\!\xi{\rm r}}^*(s)G_{\!\xi{\rm r}}(s)\\
&+\kappa F_{12}G_{\!\xi\!\sigma}(s)+\kappa G_{\!\xi\!\sigma}^*(s)F_{12}^{\T}\\
&+\kappa G_{\!\xi\!\sigma}^*(s)F_{22}G_{\!\xi\!\sigma}(s)+\kappa F_{11},\\
\Psi_{12}(s,\gamma,\kappa) &={\gamma}G_{\!\xi{\rm p}}^*(s)G_{\!a{\rm p}}(s)-G_{\!\xi{\rm r}}^*(s)G_{\!a{\rm r}}(s)\\
&+\kappa G_{\!\xi\!\sigma}^*(s)F_{22}G_{\!a\!\sigma}(s)
+\kappa F_{12}G_{\!a\!\sigma}(s),\\
\Psi_{22}(s,\gamma,\kappa) &= {\gamma}G_{\!a{\rm p}}^*(s)G_{\!a{\rm p}}(s)-G_{\!a{\rm r}}^*(s)G_{\!a{\rm r}}(s)\\
&+\kappa G_{\!a\!\sigma}^*(s)F_{22}G_{\!a\!\sigma}(s).
\end{aligned}
$$
\begin{Theorem}\label{Freq_OOG}
Assume that the matrix $A$ is Hurwitz stable, the pair $\left(A,[q\,\,B]\right)$ is controllable, and there exist scalars $\kappa\ge0$ and $\gamma>0$ such that
\begin{equation}\label{thm_freq1}
\Psi(i\omega,\gamma,\kappa)\le0 \quad \forall \omega\in\mR,
\end{equation}
and
\begin{equation}\label{thm_freq2}
{\gamma} C_{\rm p}^{\T}C_{\rm p}-C_{\rm r}^{\T}C_{\rm r}+\kappa r^{\T}F_{22}r\ge0.
\end{equation}Then 
$OOG\le\frac{1}{\gamma}$.
\end{Theorem}
\begin{proof}
Define the matrix $\bar B=[q\,\,B]$. 
By direct calculations, we obtain that \eqref{thm_freq1} leads to 
\begin{equation}\label{KYP_stat1}
\begin{bmatrix}
(i\omega I-A)^{-1}\bar B\\
 I \end{bmatrix}^*\left[\Psi_1(\gamma)+\Psi_2(\kappa)\right]\begin{bmatrix}
(i\omega I-A)^{-1}\bar B\\
 I \end{bmatrix}\le 0
\end{equation}
for all $\omega\in\mR$.
The inequality \eqref{thm_freq2} guarantees that the upper left block of the matrix $\left[\Psi_1(\gamma)+\Psi_2(\kappa)\right]$ is positive semidefinite.
Since, in addition, $A$ is Hurwitz stable and $\left(A,\bar B\right)$ is controllable, from the Kalman-Yakubovich-Popov (KYP) lemma \cite{Yak62,RANTZER19967,yakubovich2004}, we obtain
$$\left[\Psi_1(\gamma)+\Psi_2(\kappa)\right]+
\begin{bmatrix}
A^{\T}P+PA& P\bar B\\
 * & 0 \end{bmatrix}\le 0
$$
with $P\ge 0$, i.e., \eqref{thm_LMI} is satisfied. 
Therefore, the result of Theorem~\ref{Freq_OOG} follows from Theorem~\ref{LMI_OOG}. 
\end{proof}
\begin{Remark}
If the pair $\left(A,[q\,\,B]\right)$ is controllable and $\det(i\omega I-A)\ne0$, then, from the KYP lemma, it follows that the condition \eqref{thm_freq1} is necessary for the feasibility of the LMI \eqref{thm_LMI}.
\end{Remark}

\section{Numerical Example}
In this section, we will demonstrate, how nonlinear attacks can affect  the system performance.
Consider the system \eqref{sys_nonl}--\eqref{obs_lin} with the following parameters
\begin{equation}\label{sys_ex}
\begin{aligned}
A_{\rm p} & = \begin{bmatrix}
1&-2&-1\\
0&-0.5&0\\
0&0&-0.1\\
\end{bmatrix},\,\, 
B_{\rm p}= \begin{bmatrix}
0\\
1\\
1\\
\end{bmatrix},\\
C_{\rm m, o} & = \begin{bmatrix}
1&0&0\\
0&0&1\\
\end{bmatrix},\,\, 
C_{\rm p, o}= \begin{bmatrix}
0&1&0
\end{bmatrix},\,\, 
D_{\rm p, o}=0,
\end{aligned}
\end{equation}
i.e., $x=\left[x_1,x_2,x_3\right]^{\T}$, $y_{\rm p} = x_2$, $y_{\rm m} =  \left[x_1,x_3\right]^{\T}$.
Note that, without loss of generality, we take $N=0$ (no model nonlinearities) to clearly demonstrate the effect of a nonlinearity injected by the adversary. 
The controller and observer gains are chosen as
$$
L=[2.43, -3.24, -0.66],\quad K=\begin{bmatrix}
3&0&0\\-1&0&0.9\\
\end{bmatrix}^{\T},
$$
such that the eigenvalues of the matrices $A_{\rm p}+B_{\rm p}L$ and $A_{\rm p}-KC_{\rm m,o}$ are $\left\{-1,-2,-0.5\right\}$.
\begin{figure}
\centering
\includegraphics[width=3.5in]{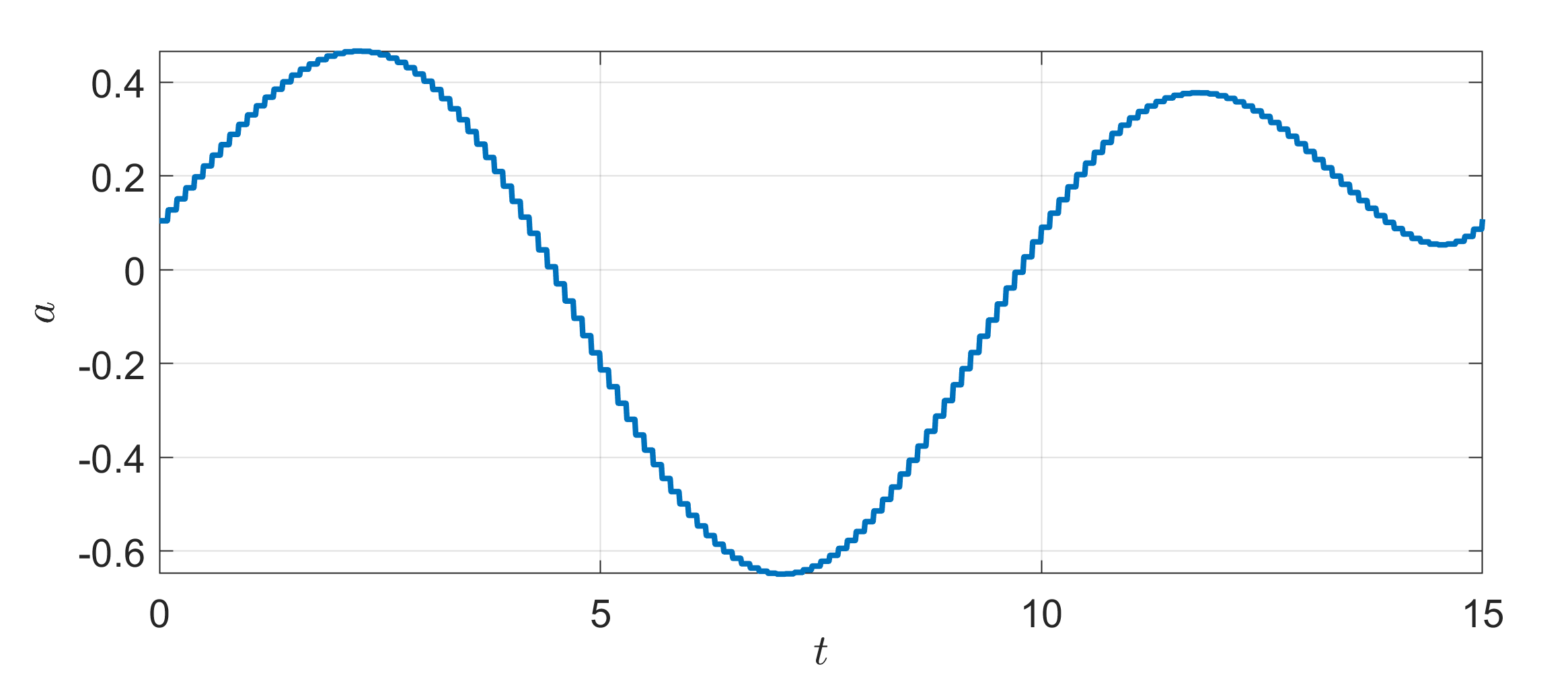}
\caption{The suboptimal injection attack signal that maximizes the output-to-output gain of the sampled-data system \eqref{sys_ex}, with a sampling step of $t_s=0.1$ and a finite time horizon of $T=15$, for $\varphi(y_{\rm m})=y_{\rm m}$.}\label{subopt_a}
\end{figure}

According to \eqref{nonl_att_output_additive}, the adversary applies a nonlinear transformation $\varphi$ and also injects an additive signal $\Gamma_{\rm y}a(t)$ into the sensor measurements.
Assume that $
\Gamma_y=\begin{bmatrix}
0\\
1\\
\end{bmatrix}$, 
i.e., the additive signal $a(t)$ can only be added to the second component, $x_3$, of the measured output $y_{\rm m}$.

\begin{figure}
\centering
\includegraphics[width=3.5in]{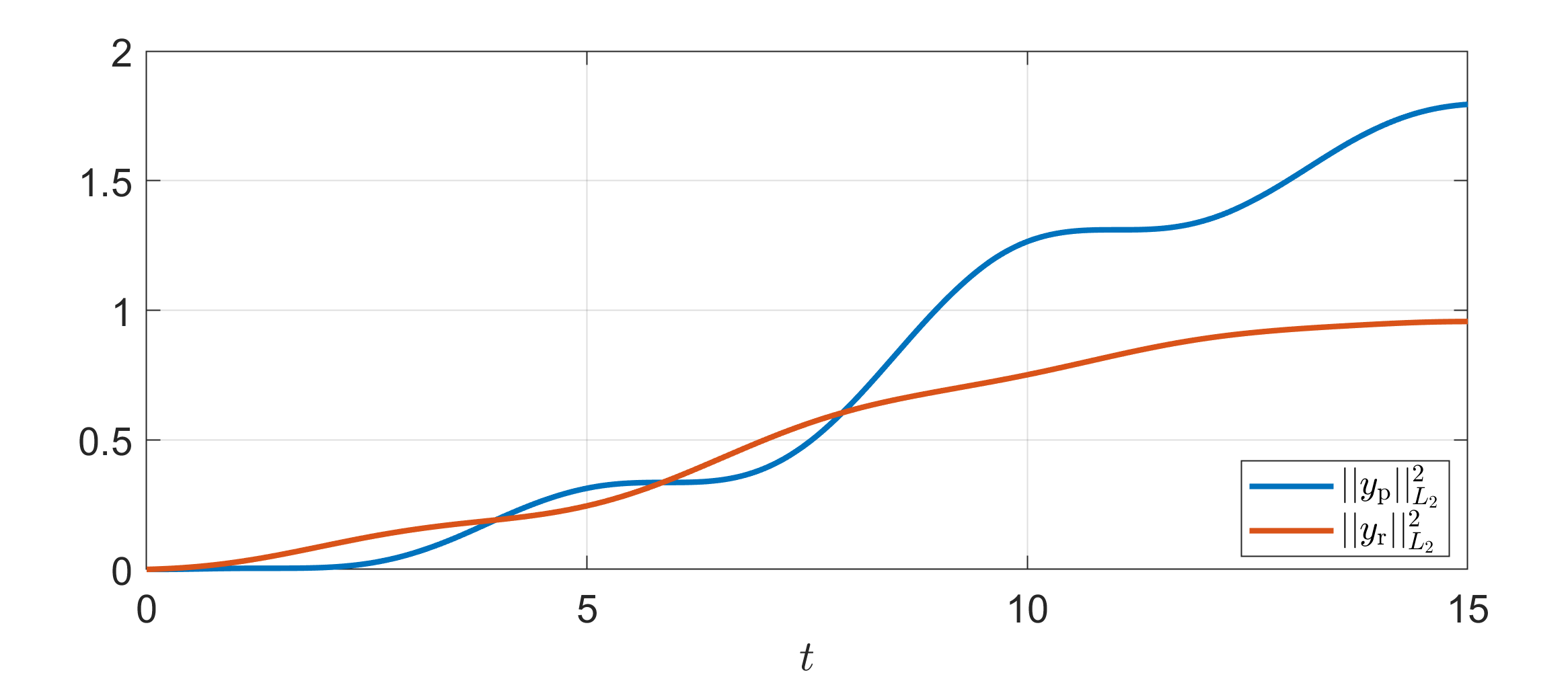}
\caption{The corresponding performance and residual output energy for the suboptimal signal $a(t)$ from Figure~\ref{subopt_a} in the linear case, yielding $OOG=1.79$.}\label{subopt_lin}
\end{figure}

In the linear case, i.e., when $\varphi(y_{\rm m})=y_{\rm m}$, the output-to-output gain can be calculated by solving the LMIs derived in \cite{Teixeira2021}, yielding $OOG=2.007$. 
For continuous-time systems with an infinite time horizon, finding the input signal $a(t)$ that achieves the maximal OOG can be complicated.
However, a suboptimal solution can be obtained by approximating the system with its sampled-data representation (we use a sampling step $t_s=0.1$ in this case) and considering a finite time horizon ($T=15$).  
Define the vectors ${\bf a}=\left[a[0]^{\T},\ldots, a[N]^{\T}\right]^{\T}$, ${\bf y_{\rm_p}}=\left[y_{\rm_p}[0]^{\T},\ldots, y_{\rm_p}[N]^{\T}\right]^{\T}$, and ${\bf y_{\rm_r}}=\left[y_{\rm_r}[0]^{\T},\ldots, y_{\rm_r}[N]^{\T}\right]^{\T}$. Then one can derive the matrices $\mathcal{T_{\rm_p}}$ and $\mathcal{T_{\rm_r}}$ such that ${\bf y_{\rm_p}}=\mathcal{T_{\rm_p}}{\bf a}$ and ${\bf y_{\rm_r}}=\mathcal{T_{\rm_r}}{\bf a}$, see \cite{Sri2021}.
Then the problem \eqref{OOG} can be approximated as 
\begin{equation}\label{OOG_appr}
\sup\limits_{{\bf a}, x(0)=0} {\bf a}^{\T}\mathcal{T_{\rm_p}}^{\T}\mathcal{T_{\rm_p}}{\bf a}, \quad \mbox{s.t.}\quad {\bf a}^{\T}\mathcal{T_{\rm_r}}^{\T}\mathcal{T_{\rm_r}}{\bf a}\le1.
\end{equation}
This problem can be rewritten as a generalized eigenvalue problem, see \cite{Teixeira2013}. Solving this yields a suboptimal input signal $a(t)={\rm ZOH}({\bf a})$ illustrated in Figure~\ref{subopt_a}, which results in $OOG=1.79$, as shown in Figure~\ref{subopt_lin}.
\begin{figure}
\centering
\includegraphics[width=3.5in]{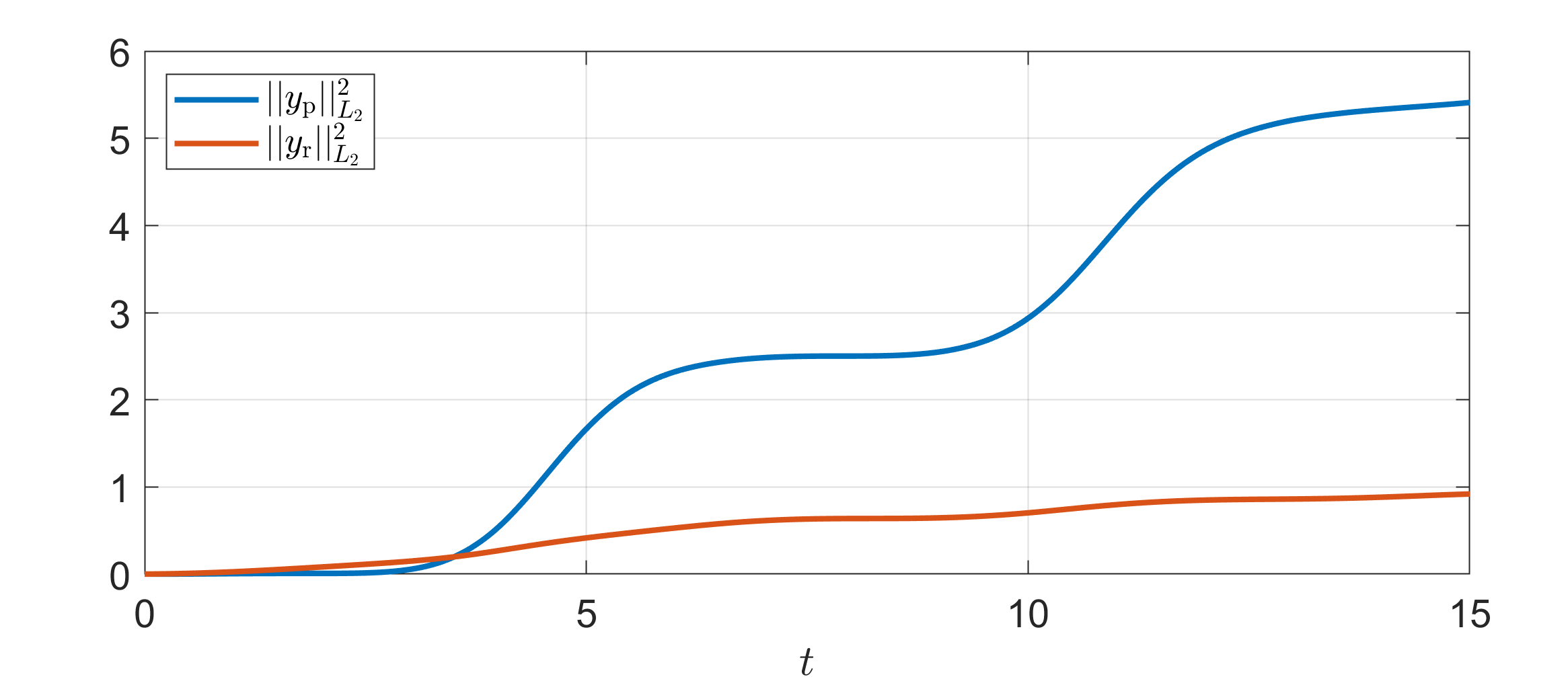}
\caption{The performance and residual output energy for the signal $a(t)$ from Figure~\ref{subopt_a} in the nonlinear case with $\varphi(y_{\rm m})=\begin{bmatrix}
x_1-0.5\sin{x_1}\\
x_3\\
\end{bmatrix}$, yielding $OOG=5.4$, which is three times larger than in the linear case.}\label{subopt_nonlin}
\end{figure}

Now, we will demonstrate that affecting the first component, $x_1$, of the measured output $y_{\rm m}$ by a nonlinear transformation can significantly increase the output-to-output gain. Assume that 
\begin{equation}\label{nonl_ex}
\varphi(y_{\rm m})=\begin{bmatrix}
x_1-0.5\sin{x_1}\\
x_3\\
\end{bmatrix}, 
\end{equation}
which belongs to the sector defined by the parameters $\mu^-=0.5$ and $\mu^+=1.1088$.
For these parameters, Theorem~\ref{LMI_OOG} provides an upper bound for the OOG of 19.1. 
It is important to note that, in the nonlinear case, identifying a specific additive signal $a(t)$ and a nonlinearity $\varphi$ from the class under consideration can be challenging.
For the transformation \eqref{nonl_ex}  and the additive attack signal depicted in Figure~\ref{subopt_a}, the OOG reaches 5.4, as illustrated in Figure~\ref{subopt_nonlin}.
Thus, performing a nonlinear attack on the $x_1$-component channel could increase the performance energy by a factor of three compared to the linear case, while the adversary remains undetected.

\section{Conclusions}
This paper addressed the analysis of nonlinear networked control systems under malicious faults and attacks, focusing on scenarios where adversaries can induce nonlinear alterations in sensor measurements. We derived frequency-domain conditions ensuring absolute stability for systems with quadratically constrained nonlinearities and extended the output-to-output gain metric to quantify performance degradation in nonlinear systems under stealthy attacks. Using dissipativity theory, computationally efficient LMI conditions and frequency-domain tests were provided to estimate upper bounds on the OOG.

Future work may explore extensions to multiplicative attacks, as well as nonlinear time-delay systems, which introduce additional challenges in both stability and performance analysis.

\bibliographystyle{IEEEtran}
\bibliography{references}



\end{document}